# Fast Krylov Space Methods for Calculation of Quark Propagator

ARTAN BORIÇI, PHILIPPE DE FORCRAND

*IPS, ETH-Zentrum*

*CH-8092 Zürich*

*borici@ips.id.ethz.ch forcrand@ips.id.ethz.ch*

## Abstract

Different recently developed Krylov space methods for solving linear systems are studied and compared for the solution of the Dirac equation on the lattice. Stabilized Biconjugate Gradient is shown to be a robust and efficient solver for the calculation of the Wilson quark propagator in lattice QCD.

## 1 Introduction

The Wilson formulation of Quantum Chromodynamics (QCD) on the lattice [1] produces the nonhermitian Dirac matrix given by

$$W_{mn} = \delta_{mn} - \kappa \sum_{\mu=1}^{4} \left[ (\mathbf{1} + \gamma_\mu) U_{\mu m} \delta_{m+\hat{\mu},n} + (\mathbf{1} - \gamma_\mu) U^\dagger_{\mu m-\hat{\mu}} \delta_{m-\hat{\mu},n} \right]$$

where $\gamma_\mu$ are four by four Dirac hermitian matrices satisfying $\gamma_\mu \gamma_\nu + \gamma_\nu \gamma_\mu = 2\delta_{\mu\nu}$ and $U_{\mu m}$ are three by three unitary, determinant one matrices residing on the site number $m$ of the four dimensional hypercubic lattice, with boundary conditions conventionally chosen periodic in the first three directions and antiperiodic in the fourth. $\kappa$ is called the hopping parameter; it takes value in the interval $[0 - \kappa_c]$ so that the eigenvalues of $W_{mn}$ are centered at unity, and circumscribed by an ellipse of axes varying from zero to one. $\kappa_c$, the value of $\kappa$ which makes $W$ singular, varies from $1/8$ to $1/4$ depending on the fluctuations of the $U_{\mu m}$'s. The size of the matrix $W$ is $12L^4 \times 12L^4$, where $L$ is the size of the hypercubic lattice; for interesting cases $L \geq 32$, so we deal with over $10^7$ equations.

We want to solve the linear systems $WG = \mathbf{1}$ for the quark propagator $G_{mn}$ in the interesting physical case when $W$ becomes near-singular during simulations of QCD. Since the matrix is huge this absorbs the bulk of the computer time devoted to the simulation. Different stationary iterative methods of Gauss-Seidel [2] type, and nonstationary iterative methods like Conjugate Gradients (CG) and Conjugate Residuals (CR) have been commonly used for solving our system [3]. However CG needs the matrix to be positive definite and hermitian and this is achieved by squaring, which slows the convergence. And because of storage constraints, CR is normally truncated to CR(1), which converges slowly too.

We present here some other class of Krylov space methods [4] with faster convergence properties. For our matrix the best method turns out to be the very recently developed BiCGstab2 [5]. In the next section we give a general idea of Krylov space methods and illustrate them for the case of our QCD Dirac matrix in the subsequent section.

## 2 Krylov space methods from Lanczos process

We solve the nonsingular system $Ax = b$ by an iterative method and start from an initial guess $x_0$ so that the initial residual error is $r_0 = b - Ax_0$. We want our $n$th residual to be in the space constructed from the set of vectors $r_0, Ar_0, ..., A^n r_0$. This is the Krylov space hence the name of the methods derived from this requirement.

The general approach of Krylov space-based methods such as CG and CR is to build an orthogonal basis for the Krylov space. The method of Conjugate Gradients for positive definite and hermitian matrices [6] makes use of short recurrences on the residuals. Indeed we can construct a Lanczos process [7] to generate an orthogonal basis for our residual vectors and call $q_0, q_1, ...$ the corresponding normalized set. Starting from $r_0$ we call $\beta^2 = <r,r>$, $\alpha = <q, Aq>$ and iterate

$$r_1 = Aq_0 - \alpha_0 q_0$$
$$r_2 = Aq_1 - \alpha_1 q_1 - \beta_1 q_0$$
$$......$$

until the process breaks down, that is a complete Krylov space is generated. Our original matrix is then similar to the symmetric tridiagonal matrix with diagonal $\alpha_0, \alpha_1, ...$ and subdiagonal $\beta_1, \beta_2, ...$ . CG is usually implemented as a set of two recurrence formulas with two terms instead of three as above, because this is more stable numerically [6]. Since our matrix is nonhermitian we cannot use short recurrences here but rather try to form a modified Gramm-Schmidt orthogonalization, maintaining the minimal property of the method for the residuals [6] at the cost of a larger storage. This leads to the set of Conjugate Residuals methods.

However we can also replace the orthogonal sequence for the residuals by two mutually orthogonal sequences if we give up the minimization property. The second sequence is updated again through a Lanczos process now on $W^\dagger$ and on some other arbitrary starting vector $\tilde{r}_0$. This is the method of Biconjugate Gradients (BiCG). The convergence of this method is rather irregular and the $LU$ decomposition of the resulting (now nonhermitian) tridiagonal system may fail; if one solves the latter system as a least squares problem we obtain the method of Quasi-Minimal Residuals (QMR). That method uses also the *look-ahead* technique to avoid the near breakdowns of the Lanczos process.

From BiCG one can derive another class of methods where the polynomial for updating the residuals is squared; in that case the norm of initial residual will be reduced twice. This gives first the Conjugate Gradient Squared method (CGS) where however the irregularity of the convergence is now more pronounced. To accelerate and stabilize the convergence, instead of squaring the polynomial, a different polynomial can be applied to the original Lanczos polynomial; it is built up in factors determined at each step by minimizing locally the residual. The BiCG stabilized (BiCGstab) method obtained in this way can be extended to the more recent BiCGstab2 where after one step of BiCGstab the residual is minimized over a two dimensional space, so that the stabilizing polynomial is a linear combination of two previous polynomials.

# 3 Behaviour of Krylov space methods for the QCD Dirac matrix

We have solved the Dirac equation for the quark propagator in two limiting cases. First in the free case, that is $U_{\mu n} = \mathbf{1}$ for all directions $\mu$ and sites $n$, i.e. in the absence of any fluctuations ("cold start") where $W$ is singular for $\kappa \equiv \kappa_c = 0.125$. Then for the completely random case ("hot start") where $U_{\mu n}$ are independent random matrices and $\kappa_c$ is 0.25. In both cases we choose $\kappa$ to be 1% less than $\kappa_c$. All the methods need two matrix–vector multiplications per iteration, except CR, where only one is needed. We display the convergence behaviour of the algorithms by giving the squared norm of the residual normalized with respect to the first residual, on a logarithmic scale, as a function of matrix–vector multiplication pairs, thus allowing for direct efficiency comparisons.

From all the methods described above we found that BiCGstab2 is more than six times faster than CR(1) for cold start and 35% faster for hot start respectively; comparing to CG similar conclusions hold with cold and hot start reversed. This is shown in Fig. 1 where an $8^4$ lattice is used. The long tail of CG in the hot start is due to the squared condition number so that this method is not recommended. In Figs. 2 and 4 we show the distribution of eigenvalues for the cold and hot start and for lattice sizes $8^4$ and $4^4$ respectively. In the free case (Fig. 2) (cold start) they are clustered. As a consequence the corresponding Lanczos process needs less steps to converge. This is not the case if fluctuations are present in the system, as shown in Fig. 4. QMR appears to be a solver with smooth convergence but less efficient than BiCGstab2. In Fig. 3 we see that the convergence of BiCG can be improved in the following way: QMR is faster and smoother; CGS is faster than BiCG and QMR, but the irregularity of BiCG is more pronounced; it is stabilized by BiCGstab2, which is at the same time the fastest and most efficient algorithm we found.

# 4 Conclusions

BiCGstab2 is found to be the winner among all methods described in section 2. This can be understood for two reasons: first BiCGstab2 applies two polynomials to the starting residual instead of one, then the very irregular convergence is stabilized by introducing a local minimal property, and this is very important because of large deviations that might occur in the solution. In this contribution we did not consider the effect of preconditioning on the original matrix, but note that we have used polynomial preconditioners or hopping parameter expansion combined with even–odd partitioning of the lattice sites. Further research should allow us to use the information of one Lanczos process to solve simultaneously the system for different right-hand sides. This will speed up calculations of hadron propagators, where 12 right-hand sides must be considered.